\documentclass[final,3p,times,twocolumn]{elsarticle}
\newcommand{\W}{7.5cm}
\usepackage{graphicx,amsmath,amssymb}
\usepackage[running]{lineno}

\begin{document}
\begin{frontmatter}
\title{Anisotropy of tracer dispersion in rough model fractures  with sheared walls.}
 \author[label1,label2]{Boschan A.}
 \author[label1]{Auradou H.}
\author[label2]{Ippolito I.}
\author[label2]{Chertcoff R.}
\author[label1]{Hulin J.P.}

 \address[label1]{Univ Pierre et Marie Curie-Paris6, Univ Paris-Sud, CNRS, F-91405.
   Lab FAST, Bat 502, Campus Univ, Orsay, F-91405, France.}
 \address[label2]{Grupo de Medios Porosos, Departemento de F\'\i sica, Facultad de Ingenier\'\i
a, Universitad de Buenos Aires, Paseo Col\'on 850, 1063
Buenos-Aires, Argentina.}

\begin{abstract}
Dispersion experiments are compared for two transparent
model fractures with identical complementary rough walls  but
with a relative shear displacement
$\vec{\delta}$  parallel ($\vec{\delta} \parallel \vec{U}$) or perpendicular
 ($\vec{\delta} \perp \vec{U}$) to the flow velocity $\vec{U}$.
 The structure of the mixing front is characterized by mapping the local normalized local transit time
 $\bar t(x,y)$  and dispersivity $\alpha(x,y)$.
  For $\vec{\delta} \perp  \vec{U}$, displacement fronts display large fingers: their geometry
  and the  distribution of $\bar t(x,y)U/x$ are  well reproduced by  assuming  parallel
 channels  of hydraulic conductance deduced from the aperture field.
For $\vec{\delta} \parallel  \vec{U}$, the front is flatter and $\alpha(x,y)$ displays a narrow
 distribution and a Taylor-like  variation with $Pe$.
\end{abstract}
\end{frontmatter}

Channelization is a key characteristic of flow and transport in
fractured rocks~(\cite{nas}) and results  frequently from the occurence of
relative shear displacements of the two fracture surfaces during
fracturation~(\cite{OlssonB93,GentierLAR97}).
Such displacements (named $\vec{\delta}$ thereafter)
 have been shown both experimentally and numerically
~(\cite{matsuki2006,Auradou06}) to create channels and
ridges perpendicular to $\vec{\delta}$.
Their length depends on
 the multiscale geometry of the fracture walls and, even
for small amplitudes $\delta$, may be a
significant fraction of the fracture size.
 The permeability is then anisotropic: both its value and the correlation length
 of the velocity field are higher for a mean flow parallel to these channels
 ({\it i.e.}  perpendicular to $\vec{\delta}$).

The objective of this communication is to demonstrate experimentally that  this type
of channelization also induces  a strong anisotropy of the
magnitude and properties of tracer dispersion. This is achieved
by comparing dispersion for mean flows parallel and perpendicular
 to the direction of the channels, but with identical flow parameters
and geometry otherwise. A previous work~(\cite{Boschan2007}) studied dispersion
in one similar model  (with a lower value of $\delta$)
but  with the different objective of analyzing the influence of the fluid rheology.
Here,
the dynamics of the process, {\it i.e.} the variation with
distance of the geometry and thickness of the mixing front
is more specifically compared in the parallel and perpendicular configurations.

Many experiments on solute spreading in fractures have
been reported: \cite{Neretnieks82, Keller2000,
Lee2003} observed dispersion coefficients $D$ increasing linearly
with the mean flow velocity $U$ ({\it i.e.} the dispersivity
$\alpha = D/U$ is  constant). However, these measurements were all
realized at the outlet of the sample with no information on the development of the
mixing front with distance. Measurements  by~\cite{Park97} used
radioactive tracers,  still with a resolution  too low to
investigate local spreading. In all these papers, the anisotropy
of dispersion is not investigated and (except for
\cite{Lee2003}) little information is available on  the relative
position of the fracture walls.

We use transparent model fractures
allowing  for high resolution optical
 concentration measurements over their  full area.
 The models are mounted vertically between a light panel and a
   $16$ bits Roper digital camera.
  Fluid flow takes place between two self-affine rectangular rough walls  of same
   characteristic exponent $H = 0.8$ as in many fractured rocks~(\cite{Poon1992}).
 The mean flow velocity $\vec{U}$ is parallel to the length $L_x = 350\, \mathrm{mm}$
 of the  walls (their width is $L_y = 90\,  \mathrm{mm}$).
 The two walls are complementary and identical in the two models  and they match perfectly when put in contact;
 then, they are pulled away
 normal to their mean surface and a lateral  shear  $\vec{\delta}$,
  parallel  or  perpendicular to  $\vec{U}$ ({\it i.e.}  to $x$) is introduced.
 In these two configurations,  referred to
  as $\vec{\delta} \parallel \vec{U}$ and $\vec{\delta} \perp \vec{U}$, the mean
  velocity $\vec{U}$ is therefore respectively perpendicular and parallel to the channels
  created by the shear.
Both $\delta$ and the mean aperture $a$ are equal to  $0.75\, \mathrm{mm}$.
The standard deviation of the
   aperture $\sigma_a = 0.144\ \mathrm{mm}$ is larger than for the similar models
   of \cite{Boschan2007}  ($\sigma_a = 0.11 \mathrm{mm}$): as a result,
    the flow field is found to be more strongly channelized.

The fluids are shear thinning $1000\ ppm$ solutions
   of scleroglucan in water with a high constant  viscosity ($\mu_0 \approx 4500\ mPa.s$) at low shear rates
   $\dot{\gamma} \le \dot{\gamma}_0$    preventing  the appearance of unwanted buoyancy driven flows~(\cite{tenchine05}).
   One has $\dot{\gamma}_0 = 0.026\, \mathrm{s^{-1}}$: for a viscous Newtonian flow between parallel plates
at a distance $a$, the corresponding mean velocity is
    $U_0 = a \dot{\gamma}_0/6 = 3 \times 10^{-3}\, \mathrm{mm/s}$.
At shear rates $\dot{\gamma} \ge  \dot{\gamma}_0$, the viscosity
    decreases as $\mu \propto \dot{\gamma}^{n-1}$ with $n = 0.26$ (see~\cite{Boschan2007}).
 One of the fluids contains $0.2 \ g/l$ of blue dye and the
 densities are matched by adding NaCl to the other.
The flow velocity $U$ is constant during each experiment with:
 $0.0024 \le U \le 0.24\ \mathrm{mm/s}$ and tracer transport is characterized
 by the dimensionless
P{\'e}clet number $Pe=Ua/D_m$  where $D_m = 6.5 \ 10^{-10}
\mathrm{m^2/s}$ is the molecular diffusion coefficient of the dye.
The experimental procedure and the determination of
dye concentration maps  from images recorded at constant time
intervals are described by~\cite{Boschan2007}.

\begin{figure}[ht]
\vskip 1cm
\center {\includegraphics[width=\W]{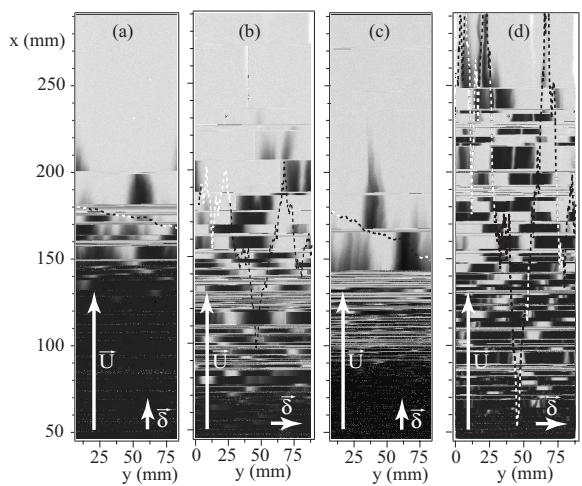}}
\vskip 1cm
\caption{Maps of the
relative concentration $c$ of the displaced fluid (white $c = 1$,
black $c = 0$)  in two transparent models. (a)-(c): $\vec{\delta} \parallel \vec{U}$;
 (b)-(d): $\vec{\delta} \perp \vec{U}$. Mean velocities: (a)-(b):
 $U = 0.0125 \ mm/s$, $Pe = 14$; (c)-(d): $U = 0.25 \ mm/s$, $Pe = 285$.
The injected volume of displacing fluid is half the void
space.}\label{fig:map}
\end{figure}

Fig.~\ref{fig:map} shows maps obtained in the two model
fractures at two different velocities $U$. If $\vec{\delta} \perp \vec{U}$,  two
fingers soar upwards with a large trough in between
(Figs.~\ref{fig:map}b and \ref{fig:map}d): they correspond to faster
paths parallel to $\vec{U}$ (Fig.~\ref{fig:map}b) and their
amplitude increases with $U$. For
$\vec{\delta} \parallel \vec{U}$, the front is smoother
(Figs.~\ref{fig:map}a and \ref{fig:map}c) while its mean slope and
the size of the indentations still increase with the velocity.

The large structures in Figs.~\ref{fig:map}b and
\ref{fig:map}d reflect velocity contrasts between the channels
created by the shear. They are well reproduced by modelling
 the fracture aperture field as a set of independent parallel
channels of aperture $a(y)=<a(x,y)>_x$~(\cite{Auradou06,Auradou08}).
A particle starting at a transverse distance
    $y$ at the inlet is assumed to move at a velocity proportional
  to $a(y)^{(n+1)/n}$ where   $n = 0.26$ for $U > U_0$ and $n = 1$  for
 $U < U_0$.

 The profile $x_f(y,t)$ of the front
at a time $t$ is then:
\begin{equation}\label{eq:xfront}
x_f(y,t)=\frac{\overline{x(t)} \
a(y)^{(n+1)/n}}{<a(y)^{(n+1)/n}>_y},
\end{equation}
where $\overline{x(t)} = <x_f(y,t)>_y$ and  $<a(y)^{(n+1)/n}>_y$ are averages
calculated over $y$.
The profiles  computed using  Eq.~(\ref{eq:xfront}) and  the actual aperture fields  appear in
Figs.~\ref{fig:map}a-d as dotted lines (from the above
discussion, one assumes that $n = 1$ at
the lowest velocity (Figs.~\ref{fig:map}a and \ref{fig:map}b) and
$n = 0.26$ at the highest one (Figs.~\ref{fig:map}c and
\ref{fig:map}d)).
Eq.~(\ref{eq:xfront}) predicts well the location and shape of the ``fingers'' and ``troughs''
at both velocities for $\vec{\delta} \perp \vec{U}$
although their amplitude is slightly  underestimated
in Fig.~\ref{fig:map}b. In this latter case, one has $U \sim U_0$, corresponding
to a transition regime between the power law and Newtonian rheologies.

These results demonstrate that, for $\vec{\delta} \perp \vec{U}$
({\it i.e.} if $\vec{U}$ is parallel to the channels created by the shear),
 the large scale features of solute transport are determined
by the contrasts between the velocities in these channels which
increase with their aperture. Then,
front spreading is purely convective and the total width
$\Delta x$ of the front parallel to $\vec{U}$ ({\it i.e.} the distance between
the tips of the fingers and the bottom of the troughs) increases linearly
with distance as $x\ \Delta U/U$ ($\Delta U/U$ = typical relative
velocity contrast between the different channels). These curves
also demonstrate  that the difference between the sizes of the fingers in
the two cases are accounted for by the different rheological
behavior of the fluids: the velocity contrasts (and therefore the
size) are amplified for $Pe = 285$ (shear-thinning
power law domain) compared to the vicinity of
the Newtonian constant viscosity regime  ($Pe = 14$).

For  $\vec{\delta} \parallel \vec{U}$,
the features of the front  are also visible  at the same transverse distances $y$
in Figs.~\ref{fig:map}a and  \ref{fig:map}c: they reflect again a convective
 spreading of the front  due to velocity contrasts between the  flow paths.
However, in contrast with the previous case $\vec{\delta} \perp \vec{U}$,
 these features (except for the small  global slope of the front) are not reproduced
 by the theoretical model (dotted line): this was to be
expected since its underlying hypothesis  are not satisfied for $\vec{\delta} \parallel \vec{U}$.
\begin{figure}[ht]
\vskip 1cm
\center {\includegraphics[width=\W]{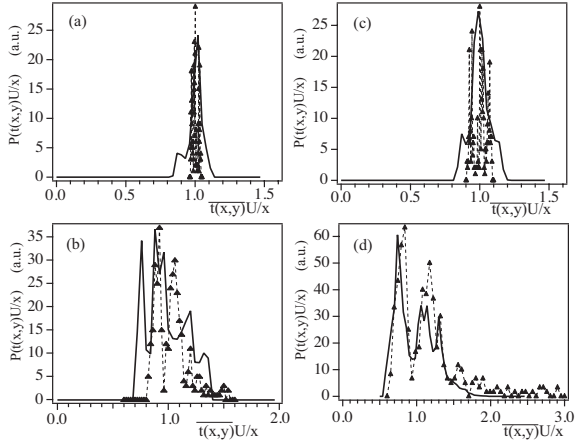}}
\vskip 1cm
 \caption{Histograms
of the experimental normalized local transit  time
$\overline{t(x,y)}U/x$ (continuous lines) for the same models
and $Pe$ values as in Figs.~\ref{fig:map} (the letters
corresponding to the different experiments are the same in both
graphs). The distribution of $\overline{t(x,y)}U/x$ has been
computed in  the upper fifth of the length of the model to make
 meaningful comparisons with the distribution of the
theoretical transit times (($\blacktriangle$) symbols and dotted
lines) determined from Eq.~(\ref{eq:xfront}).}\label{fig:distrit}
\end{figure}
The time variations of the local concentration $c(x,y,t)$  on
individual pixels complement the above study by providing
 quantitative information on the
interplay of convective and diffusive mechanisms.
 For all the experiments, $c(x,y,t)$ is found to be well fitted by
  solutions of a classical
$1D$ convection-dispersion equation for a step-like initial
variation of $c$ at the inlet ($x = 0$):
\begin{equation}
c(x,y,t) = \frac{1}{2} (1 + erf \frac{t-\overline{t(x,y)}}{\sqrt{4 D(x,y) t}}),
\label{eq:cdif}
\end{equation}
Here, $\overline{t(x,y)}$ and $D(x,y)$ are the mean transit time and
apparent dispersion  coefficient deduced from the time variation
of the relative concentration $c(x,y,t)$ of the displaced fluid at  point $(x,y)$
(transverse dispersion is  neglected here).
It will be shown below that the two
parameters provide complementary information:
$D(x,y)$ (or rather the dispersivity $\alpha(x,y) = D(x,y)/U$) characterizes the local
thickness of the mixing front while $\overline{t(x,y)}$ is  related to its global geometry.

For each experiment,  $\overline{t(x,y)}$ and $D(x,y)$  are
determined for all pixels inside the field of view.
Fig.~\ref{fig:distrit} compares experimental and theoretical probability
distributions of the normalized times  $\overline{t(x,y)} U/x$ for the same
experiments as Fig.~\ref{fig:map}.
The theoretical distribution is obtained by taking $\overline{t(x,y)} U/x = x_f(y,t)/\overline{x(t)}$,
computing the ratio $x_f(y,t)/\overline{x(t)}$ by means of   Eq.~(\ref{eq:xfront})
for all $y$ values and determining finally the distribution of the   results.
As expected,
the distributions are much broader for $\vec{\delta} \perp \vec{U}$
 (Figs.~\ref{fig:distrit}b-d) than for $\vec{\delta} \parallel \vec{U}$
(Figs.~\ref{fig:distrit}a-c). For $Pe = 285$,
the  experimental distribution for flow parallel to the channels
coincides very well with the theoretical one and displays two
peaks reflecting the structuration of the flow. At $Pe = 14$, the
width and global shape of the experimental and theoretical
distributions  are overall similar and narrower than
for $Pe = 285$ due to the lower velocity contrast
in the Newtonian limit. For $\vec{\delta} \parallel \vec{U}$
 (Figs.~\ref{fig:distrit}a-c), the distribution at
both P\'eclet numbers is much narrower than for $\vec{\delta} \perp \vec{U}$. The
mean peak corresponds to  $\overline{t(x,y)} U/x \simeq 1$;
its width increases with $Pe$ (again likely due to an
increase of the velocity contrasts) and is similar to that of the
theoretical distributions. The experimental distribution displays
however additional ``aisles'': these reflect likely
complex paths deviating from straight trajectories
parallel to $\vec{U}$.

\begin{figure}[ht]
\vskip 1cm
\center {\includegraphics[width=\W]{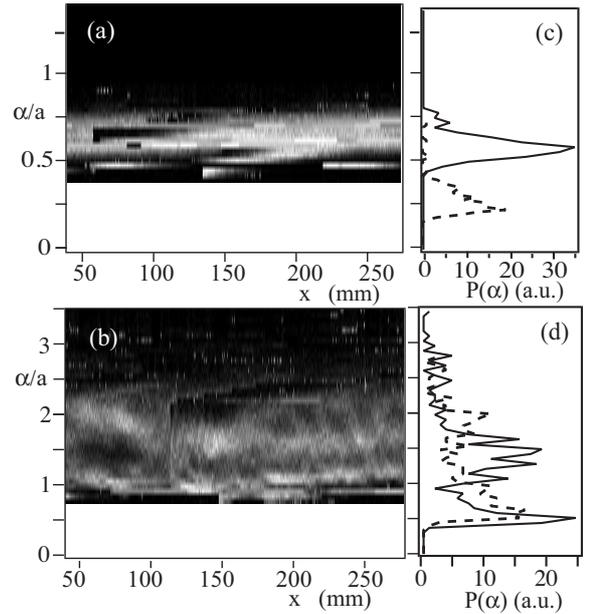}}
\vskip 1cm
\caption{(a)-(b): Histograms (grey levels) of the values of the normalized local dispersivity
$\alpha(x,y)/a$ (vertical scale) as a
 function of the distance $x$ (horizontal scale) for $Pe = 285$.
 White: maximum probability, black :
 zero probability. (c), (d): Histograms obtained at $x = 240\, \mathrm{mm}$ for $Pe = 285$
 (continuous  line) and $Pe = 14$ (dashed line).
  (a), (c): $\vec{\delta} \parallel \vec{U}$; (b), (d): $\vec{\delta} \perp \vec{U}$.}\label{fig:histold}
\end{figure}
While, from the above results, the overall geometry of the mixing front seems
to be determined mainly by convective effects, we examine now
the relative influence of convection and diffusion on the local width
of this front:  this may be
characterized by the variation of the local dispersivity $\alpha(x,y)$ with $Pe$.
Figs.~\ref{fig:histold}a-b display, for each value of $x$ (horizontal scale), the histogram
(coded in grey levels) of the corresponding values of  $\alpha(x,y)/a$ (vertical scale).

\begin{figure}[ht]
\vskip 1cm
\center {\includegraphics[width=\W]{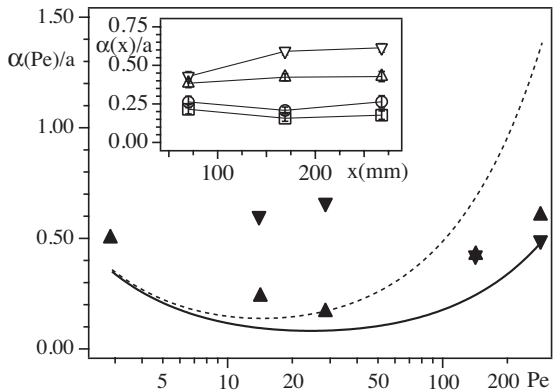}}
\vskip 1cm
\caption{Variation of the normalized dispersivity $\alpha/a$ as a function of $Pe$.
 ($\blacktriangle$)  $\vec{\delta} \parallel \vec{U}$: mean value of $\alpha/a$.
($\blacktriangledown$)   $\vec{\delta} \perp \vec{U}$: value corresponding to the first peak
 in the distribution of $\alpha/a$.
Continuous and dashed  lines:  Taylor dispersion between parallel plates
respectively  for a power law fluid of exponent $n = 0.26$ and a Newtonian fluid ($n = 1$).
 Inset: variation of the mean value $<\alpha(x,y)>_y$ as a function of
the distance $x$ for $\vec{\delta} \parallel \vec{U}$;
 $Pe= 285$ ($\triangledown$), $Pe= 142$  ($\triangle$), $Pe = 28.5$ ($\square$),
$Pe = 14$ ($\circ$). }
\label{fig:ldvsx}
\end{figure}

For $\vec{\delta} \parallel \vec{U}$, the probability distribution
of $\alpha(x,y)/a$ is narrow, particularly at high flow velocities
(Figs.~\ref{fig:histold}a and c). Moreover,  the mean value varies little with the distance $x$ and
reaches  a  constant value   $\alpha(Pe)/a$ for $x \ge 100 \mathrm{mm}$ (see inset of Fig.~\ref{fig:ldvsx}).
The increase with time of the local front
thickness  is therefore  diffusive  and can be characterized by a single dispersivity value $\alpha(Pe)/a$.

For $\vec{\delta} \perp \vec{U}$ (Figs.~\ref{fig:histold}b and d), the distribution of the values of $\alpha/a$ is
much broader and displays a ``tail'' at large values of $\alpha/a$. At the two highest velocities, the distribution
displays two peaks (solid curve in  Fig.~\ref{fig:histold}).
While the value of $\alpha(x)/a$ corresponding to the peak(s) seems to reach a limit at long distances $x$,
the global width of the distribution keeps increasing with  $x$. In contrast to the case  $\vec{\delta} \parallel \vec{U}$,
 the increase of the local thickness of the mixing front with $x$ is not simply  diffusive and cannot be characterized
 by a single dispersivity parameter.

The variation of $\alpha(Pe)$ with $Pe$ for $\vec{\delta} \parallel \vec{U}$
 displayed in Fig.~\ref{fig:ldvsx}
provides quantitative information on the local dispersion mechanisms: the values are
 similar (although slightly higher) to the predictions for Taylor dispersion
with a power law fluid (continuous line). At low $Pe$'s, the values obtained for $n = 1$ (dashed line)
and for $n = 0.26$ are similar and the transition towards a Newtonian behaviour should not
influence the variations.
 A dominant contribution of Taylor dispersion has already been
demonstrated in models with a randomly distributed aperture of short correlation
length~(\cite{Detwiler00}); however, at low $Pe$'s,
 an additional  geometrical dispersion regime  ($\alpha=cst(Pe)$) was observed
and is not present here.

In the opposite case ($\vec{\delta} \perp \vec{U}$), no single value of $\alpha$
characterizes dispersion in the whole model. However, the first peak in
the distribution of $\alpha$ likely reflects dispersion in regions of low transverse  gradient
of the mean front velocity in the plane of the model: there, additional spreading due to transverse diffusion
in the velocity gradient should be reduced.
The values of $\alpha/a$ corresponding to this first peak
have therefore been plotted in Fig~\ref{fig:ldvsx} ($\blacktriangledown$) as a function of $Pe$.
At high velocities, they are indeed close to those corresponding to Taylor dispersion (and to the
other model); at low velocities, they remain higher.

Overall, the present experiments  demonstrate  that solute dispersion in a
 channelized rough fracture depends crucially of the orientation  of the flow.
Optical measurements allowed  us to characterize
the dynamics of dispersion at both the global and local scales from
mean transit times and local dispersivities deduced from local concentration
variations.

For  $\vec{\delta} \perp \vec{U}$ ($\vec{U}$ parallel to the channelization),  the
large scale geometry of the displacement front is controlled  by the velocity
contrasts between the channels. At all $Pe$ values
 ($14 \le Pe \le 285$) the geometries of the large fingers
and troughs in the front and the distribution of the local transit times $\overline{t(x,y)} U/x$
are  well predicted  from a transverse effective permeability profile
computed using  the aperture field.
Moreover, differences between the finger sizes  at low and high $Pe$'s are explained
by variations of the fluid rheology (shear-thinning at high Pe's and Newtonian at low ones).
These results confirm fully the convective origin of the large scale
structures of the front and the relevance of local  measurements all along the flow paths:  measuring
 only  variations of the mean concentration at the
outlet (as is often done practically) and fitting it to a solution of a convection-diffusion
equation might indeed lead to an incorrect identification of such processes as geometrical dispersion.

For $\vec{\delta} \parallel \vec{U}$ ($\vec{U}$ perpendicular to the channelization), the front is flatter and the distribution
of the transit times is narrower than for $\vec{\delta} \perp \vec{U}$; this reflects a
more effective sampling of the velocity  heterogeneities by the solute particles.
As could be expected, the remaining geometrical features of the front cannot
be predicted by the channel model: however they
 are still observed at the same transverse locations
at all velocities and remain of similar amplitudes (taking into account the
 variations of the rheology). This suggests that these features are again of convective
 origin.

Still for $\vec{\delta} \parallel \vec{U}$, the variation of the  local front thickness
 is diffusive and well characterized by
a single mean dispersivity $\alpha(Pe)$; its  dependence  on $Pe$
suggests a dominant influence of Taylor dispersion.
For $\vec{\delta} \perp \vec{U}$, in contrast, the broad distribution of
$\alpha(x,y)$ does not allow one to define a meaningful global dispersivity. In addition,
even the lowest values of $\alpha(x,y)$ (corresponding to simple
flow paths) are larger than those expected for Taylor dispersion except at
the highest $Pe$ values; values corresponding to Taylor dispersion are also expected
at smaller values of $\delta$ leading to a reduced disorder of the flow field~\cite{Boschan2007}.
This latter difference, as well as the tail in the distribution,  reflect
the increasing influence of transverse molecular diffusion inducing tracer exchange
with adjacent flow paths of different velocities and enhancing dispersion.

This set of results is highly relevant to the interpretation of
 field observations~(\cite{Becker2000}).
However, the influence of the length of the samples is an important issue and will
need to be investigated before transposing these results.
In particular, although the front geometries are determined in both configurations
by spatial variations of the flow velocities, a geometrical dispersion regime has
never been observed, in contrast with experiments on rough fractures with
a small correlation length of the aperture~(\cite{Detwiler00}).
For much longer fractures,  transverse diffusion might be large enough for
solute particles to sample the whole  distribution of local velocities and reach
such a global diffusive spreading regime.
In the case of a broad distribution of the
hydraulic conductivities, one might also observe instead an anomalous dispersion regime as suggested
by~\cite{Bouchaud87} and \cite{Dentz2008}.
Because of the specific correlations  of the flow
velocity field for self-affine wall geometries~(\cite{Auradou06}), the corresponding exponent
would then likely depend on the characteristic roughness exponent of the fracture walls.
\section{Acknowledgments}
We are indebted to R. Pidoux for his assistance. HA and JPH are
supported by CNRS through the GdR No. $2990$) and by the EHDRA
(European Hot Dry Rock Association). This work was greatly
facilitated by a CNRS-Conicet Collaborative Research Grant and by
the Ecos Sud $A03E02$ program.

\end{document}